\documentclass[intlimits,twoside,a4paper]{article}
\usepackage{amsmath,amssymb}
\usepackage{graphicx}
\usepackage{wrapfig}
\usepackage[T2A]{fontenc}
\usepackage[cp1251]{inputenc}
\usepackage{cmpj}

\issue{2011}{14}{2}{23704}

\doinumber{10.5488/CMP.14.23704}

\title[Active conductivity of plane two-barrier resonance tunnel
structure]%
{Active conductivity of plane two-barrier resonance tunnel
structure as operating element of quantum cascade laser or
detector}
\author[M.V. Tkach \textsl{et al.}]
{M.V. Tkach\thanks{E-mail: ktf@chnu.edu.ua}\,, Ju.O. Seti, V.O.
Matijek, O.M. Voitsekhivska}
\address{Fedkovych Chernivtsi National University, 2~Kotsyubinsky~Str.,
58012~Chernivtsi, Ukraine}

\date{Received March 4, 2011, in final form May 12, 2011}

\authorcopyright{M.V. Tkach, Ju.O. Seti, V.O. Matijek, O.M. Voitsekhivska, 2011}
\begin{document}

\maketitle

\begin{abstract}
Within the model of rectangular potentials and different effective
masses of electrons in different elements of plane two-barrier
resonance tunnel structure there is developed a theory of spectral
parameters of \linebreak quasi-stationary states and active conductivity for
the case of mono-energetic electronic current interacting with
electromagnetic field.
It is shown that the two-barrier resonance tunnel structure can be
utilized as a \linebreak separate or active element of quantum cascade laser
or detector. For the experimentally studied
\linebreak
In$_{0.53}$Ga$_{0.47}$As/In$_{0.52}$Al$_{0.48}$As nano-system it
is established that the two-barrier resonance tunnel structure, in
detector and laser regimes, optimally operates (with the
biggest conductivity at the smallest exciting current) at the
quantum transitions between the lowest quasi-stationary states.

\pacs 73.21.Fg, 73.90.+f, 72.30.+q, 73.63.Hs
\end{abstract}

\section{Introduction}

During the last decades, after the creation of the first
nano-lasers by Faist and Capasso~\cite{1,2} working at the
transitions between electronic levels of size-quantization, the
evident success was achieved in the improvement of quantum cascade
lasers (QCLs)~\cite{3,4,5,6} and quantum cascade detectors
(QCDs)~\cite{7,8,9,10} with various geometric design. These
devices operate effectively in the actual terahertz range
of electromagnetic waves with the frequencies getting into the
known atmosphere transparency windows. Thus, the QCLs and QCDs are
constantly in the field of researchers' vision.

The main purpose of investigations is to optimize the parameters of nano-devices
which is actually  a hard task due to the absence of a consequent and complete
theory of physical processes in open nano-systems. As far as the active
working element in experimentally produced QCL or QCD were the open
resonance tunnel structures (RTS), with different number of barriers and
wells, the main theoretical attention was paid to the study of static and
dynamic conductivities in such nano-systems because they determine the main QCL
or QCD parameters, such as region and width of operating frequency range,
radiation intensity, exciting current and so on.

The theory of dynamic conductivity~\cite{11,12,13,14,15,16,17} of electrons
 in open RTS, as separate
active element of QCL working in ballistic regime, is developed
within the analytic solution of complete Schrodinger equation
using the simplified model of electron constant effective mass in
all parts of nano-system and $\delta$-like approximation of
rectangular potential barriers. In the cited and other papers,
based on the simplified model of RTS, important results were obtained
explaining the general properties of conductivity in open
systems, but due to the rather rough approximating
models, as it was established in reference~\cite{18}, the obtained
magnitudes of resonance energies and width of electron
quasi-stationary states in RTS (essentially determining the
conductivity magnitude) were  manifestly overestimated compared to the
more realistic model of rectangular potentials. Therefore, in the
approximated model, the problem of the optimization of QCL, QCD or
the operation of separate elements was not observed at all.

In the paper, using the model of rectangular potential wells and
barriers and considering different electron effective masses in
different parts, there is developed a theory of active conductivity
of open plane two-barrier RTS as separate nano-detector or
nano-laser. For the nano-system with In$_{0.53}$Ga$_{0.47}$As
wells and In$_{0.52}$Al$_{0.48}$As barriers  the
best geometric design of two-barrier RTS is established providing its optimal
work as  an active element of nano-detector or nano-laser, i.e.,
providing the maximal active conductivity through the RTS at the
minimal life times of electrons in the operating quasi-stationary states
(QSSs). Besides, it is shown that  contrary to the laser
where the radiation transitions between two lowest electron QSS-s
are not always optimal, the energy of mono-energetic electron beam
falling at two-barrier RTS should correspond to the energy of the lowest
QSS (from which the quantum transitions into the second resonance
electron state occur with the absorption of electromagnetic
energy) for the  detector to operate effectively.

\section{Active conductivity of two-barrier RTS}

\begin{figure}[h]
\centerline{\includegraphics[width=0.45\textwidth]{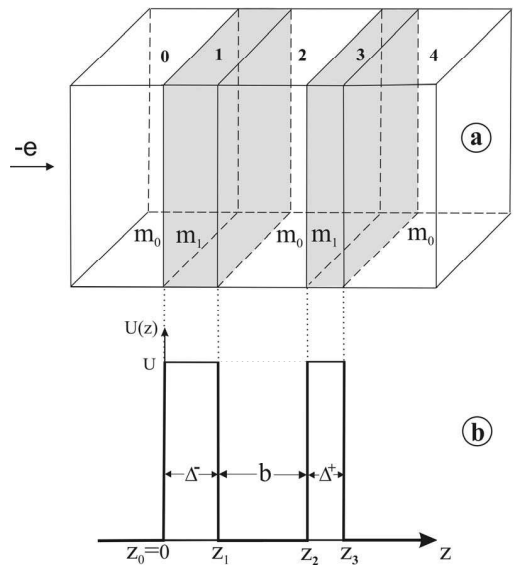}}
\caption{Geometric (a) and potential energy (b) schemes of
two-barrier resonance tunnel structure.} \label{Fig1}
\end{figure}

In Cartesian coordinate system, the open two-barrier RTS is
observed with geometric parameters shown in figure~\ref{Fig1}. The small
differences of lattice constants of RTS barriers and wells make it possible
 to study the nano-system within the models of effective masses and
rectangular potentials
\begin{equation}
 \label{eq1} m(z) = {\left\{ {{\begin{array}{*{20}c}
 {m_{0}}\,,  \hfill \\
 {m_{1}}\,,  \hfill \\
\end{array}} } \right.}
\qquad  U(z) = {\left\{ {{\begin{array}{*{20}c}
 {0, \qquad reg.\ 0,\,2,\,4, } \hfill \\
 {U, \qquad reg.\,\,1 ,\,3.} \hfill \\
\end{array}} } \right.}
\end{equation}

The electronic beam moving in the direction perpendicular to the two-barrier RTS plane,
impinges on it from the left hand side.  The electrons with energy $(E)$ and
concentration $(n_0)$ are assumed to be uncoupled. In order to obtain the
conductivity of a nano-system, determined by the density of
current flowing through it, according to the quantum mechanics,
one has to find the  wave function of  the electron dependent on time,
 interacting with the electromagnetic field periodic in time.

In the problem under study, the movement of electrons is observed as
one-dimensional ($k_{\vert \vert}=0$). Consequently, $\Psi(z,t)$
wave function satisfies the complete Schrodinger equation
\begin{equation}\label{eq2.2}\ri\hbar{\frac{{\partial\Psi(z,t)}}{{\partial
t}}}=\left[{H + H(z,t)}\right]\Psi(z,t), \end{equation} where
\begin{equation}
\label{eq2.3}H=-{\frac{{\hbar^{2}}}{{2}}}{\frac{{\partial}}{{\partial
z}}}{\frac{{1}}{{m\left({z} \right)}}}{\frac{{\partial}
}{{\partial z}}}+U(z)\end{equation} is the electron Hamiltonian in
stationary case,
\begin{equation}
\label{eq2.4}H(z,t)=-e \epsilon {\rm} \left\{
{z\left[{\theta\left({z}
\right)-\theta\left({z-z_{3}}\right)}\right]
+z_{3}}{\theta\left({z-z_{3}}\right)}\right\}\left({\re^{\ri\omega
t}+\re^{-\ri\omega t}} \right).\end{equation}

 The Hamiltonian of
an electron interacts with time  varying electromagnetic field
with frequency ($\omega $) and amplitude of electric field
intensity ($\epsilon$).

The solution of equation~(\ref{eq2.2}) in the approximation of
small signal~\cite{11,12,13,14,15,16,17} is written as follows:
\begin{equation}
\label{eq2.5}\Psi\left({z,t}\right)=\Psi_{0}\left({z} \right)\re^{
- \ri\omega_{0}t}+ \Psi_{1}\left({z,t}\right),\qquad
(\omega_{0}=E/\hbar)
\end{equation}
where $\Psi_{0}(z)$ function is the solution of stationary
Schrodinger equation
\begin{equation}
\label{eq2.6}H\Psi_{0}(z)=E\Psi_{0}(z).
\end{equation}

The first order correction in one-mode approximation is found as
\begin{equation}
\label{eq2.7}\Psi_{1}\left({z,t}\right)=\Psi
_{+1}\left({z}\right)\re^{-\ri\left({\omega
_{0}+\omega}\right)t}+\Psi_{-1}\left({z}\right)\re^{-\ri\left({\omega
_{0}-\omega}\right)t}.\end{equation}

Preserving the magnitudes of the first smallness order and taking
into account formulas~\eqref{eq2.5}, \eqref{eq2.6}, \eqref{eq2.2},
 the equation is obtained for the both parts $\Psi_{\pm1}(z)$ of
function $\Psi_{1}(z,t)$
\begin{equation}
\label{eq2.8}\left({-{\frac{{\hbar^{2}}}{{2}}}{\frac{{\partial}
}{{\partial
z}}}{\frac{{1}}{{m\left({z}\right)}}}{\frac{{\partial}}{{\partial
z}}} + U(z)-\hbar
\left({\omega_{0}\pm\omega}\right)}\right)\Psi_{\pm
1}\left({z}\right)+H\left({z}\right)\Psi_{0} \left({z}\right)=0,
\end{equation}
where
\[
H(z)=- e \epsilon \left\{ {z\left[{\theta \left({z}\right)-\theta
\left({z-z_{3}}\right)}\right]+z_{3}\theta\left({z-z_{3}}\right)}
\right\}.
\]

The solution of stationary Schrodinger problem,
equation~\eqref{eq2.6} is written as
\begin{eqnarray}
\label{eq2.9}
 \Psi _{0} (z) &=& \Psi _{0}^{(0)} (z)\theta ( - z) + \sum\limits_{p = 1}^{3}
{\Psi _{0}^{(p)} (z){\left[ {\theta (z - z_{p - 1} ) - \theta (z -
z_{p} )} \right]}}  + \Psi _{0}^{(4)} (z)\theta (z - z_{3} )
\nonumber\\
 &=& \left( {\re^{\ri k^{(0)}z} + B^{(0)}\re^{ - \ri k^{(0)}z}} \right){\rm} \theta ( -
z) + A^{(4)}\re^{ \ri k^{(4)}z}\theta (z - z_{3} ) \nonumber \\
 &&{}+ {\sum\limits_{p = 1}^{3} {\left( {A^{(p)}\re^{\ri k^{(p)}z} +
B^{(p)}\re^{ - \ri k^{(p)} z}} \right){\left[ {\theta (z - z_{p -
1} ) - \theta (z - z_{p} )} \right]}}} ,
\end{eqnarray}
where
\[
k^{(0)} = k^{(2)} = k^{(4)} = k = \hbar ^{ - 1}\sqrt {2m_{0} E} \,
, \qquad k^{(1)} = k^{(3)} = \hbar ^{ - 1}\sqrt {2m_{1} (E - U)}\,
,
\]
\[
z_{0} = 0, \qquad z_{1} = \Delta _{1}^{ -}, \qquad z_{2} = b +
\Delta _{1}^{ -}, \qquad z_{3} = b + \Delta, \qquad \Delta =
\Delta _{1}^{ -}  + \Delta _{1}^{ +}.
\]

The unknown coefficients $B^{(0)}$, $A^{(4)}$, $A^{(p)}$,
$B^{(p)}$ $(p = 1,\ 2,\ 3)$ are fixed by the conditions of wave
functions and their densities of currents continuity at all
nano-system interfaces
\begin{equation}
\label{eq2.10} \Psi _{0}^{(p)} (z_{p} ) = \Psi _{0}^{(p + 1)}
(z_{p} ), \qquad {\frac{{1}}{{m_{0(1)}} }}{\left. {{\frac{{\rd
\Psi _{0}^{(p)}} }{{\rd z}}}} \right|}_{z = z_{p}} =
{\frac{{1}}{{m_{1(0)} }}}{\left. {{\frac{{\rd \Psi _{0}^{(p + 1)}
(z)}}{{\rd z}}}} \right|}_{z = z_{p}} \qquad  (p = 0, \ 1,\ 2,\ 3
)
\end{equation}
as well as because the nano-system is an open one, from the
normalizing condition for the wave functions (at fixed $k_{\vert
\vert}  = 0$)
\begin{equation}
\label{eq2.11} {\int\limits_{ - \infty} ^{\infty}  {\Psi _{0}^{ *}
({k}'z)\Psi _{0}^{} (kz)}} {\rm} \rd z = \delta (k - {k}').
\end{equation}

The obtained wave function $\Psi _{0}$ defines the density of
electronic current and, thus, the permeability coefficient of the
system as function of energy. It is well known~\cite{18}, that
the permeability coefficient in the vicinity of their maxima is of
a quasi-Lorentz shape. Consequently, the position of maximum in
the energy scale defines the resonance energy ($E_{n}$) and the width
of Lorentz curve at half of its height fixes the resonance
width ($\Gamma_{n}$) of the corresponding QSS.

The solutions of inhomogeneous equations~\eqref{eq2.8} are the
super-positions of functions
\begin{equation}
\label{eq2.12} \Psi _{\pm 1} \left( {z} \right) = \Psi _{\pm}
\left( {z} \right) + \Phi _{\pm}  \left( {z} \right),
\end{equation}
where $\Psi _{\pm} \left({z}\right)$ are the solutions of
homogeneous and $\Phi _{\pm}\left( {z} \right)$ are solutions of
inhomogeneous equations~\eqref{eq2.8}.

The solutions of homogeneous equations \eqref{eq2.8} are found as
\begin{eqnarray}
\label{eq2.13}
 \Psi _{\pm}  (z) &=& \Psi _{\pm} ^{(0)} (z)\theta ( - z) + {\sum\limits_{p =
1}^{3} {\Psi _{\pm} ^{(p)} (z){\left[ {\theta (z - z_{p - 1} ) -
\theta (z -
z_{p} )} \right]}}}  + \Psi _{\pm} ^{(4)} (z)\theta (z - z_{3} )  \nonumber\\
& = & B_{\pm} ^{(0)} \re^{ - \ri k_{\pm} ^{(0)} z}\theta ( - z) +
A_{\pm} ^{(4)}
\re^{\ri k_{\pm} ^{(4)} \left( {z - z_{3}}  \right)}\theta (z - z_{3} )\nonumber \\
&&{} + {\sum\limits_{p = 1}^{3} {{\left[ {B_{\pm} ^{(p)} \re^{ -
\ri k_{\pm} ^{(p)} (z - z_{p - 1} )} + A_{\pm} ^{(p)} \re^{\ri
k_{\pm} ^{(p)} (z - z_{p - 1} )}} \right]}}} {\left[ {\theta (z -
z_{p - 1} ) - \theta (z - z_{p} )} \right]},
\end{eqnarray}
where
\begin{equation}
\label{eq2.14} k_{\pm} ^{(0)} = k_{\pm} ^{(2)} = k_{\pm} ^{(4)} =
\hbar ^{ - 1}\sqrt {2m_{0} (E\pm \hbar \omega )}\,, \qquad k_{\pm}
^{(1)} = k_{\pm} ^{(3)} = \hbar ^{ - 1}\sqrt {2m_{1} \left[ {(E -
U)\pm \hbar \omega}  \right]} \,.
\end{equation}

Exact partial solutions of equations~\eqref{eq2.8} are known
\begin{eqnarray}
\label{eq2.15}
 \Phi _{\pm}  (z) &=& {\sum\limits_{p = 1}^{3} {{\left[
{\mp {\displaystyle \frac{{e \epsilon}}{{\hbar \omega} }}z\,\Psi
_{0}^{(p)} \left( {z} \right) + {\displaystyle \frac{{e
\epsilon}}{{m_{p} \omega ^{2}}}}{\displaystyle \frac{{\rd\Psi
_{0}^{(p)} \left( {z} \right)}}{{\rd z}}}} \right]}}}{\left[
{\theta (z - z_{p - 1} ) - \theta (z - z_{p} )} \right]} \nonumber \\
&&{} \mp  \displaystyle {\frac{{e \epsilon}}{{\hbar \omega} }}z_{3}
\Psi _{0}^{(4)} \left( {z_{3}} \right)\theta (z - z_{3} ){\rm} .
\end{eqnarray}

Thus, the general solutions of these equations can be written as

\begin{equation}
\label{eq2.16} \Psi _{\pm 1} \left( {z} \right) = \Psi _{\pm
1}^{(0)} \left( {z} \right){\rm} \theta ( - z) + {\sum\limits_{p =
1}^{3} {\Psi _{\pm 1}^{(p)} \left( {z} \right){\rm} {\left[
{\theta (z - z_{p - 1} ) - \theta (z - z_{p} )} \right]}}} + \Psi
_{\pm 1}^{(4)}  \left( {z} \right){\rm} \theta (z - z_{3} ).
\end{equation}

The conditions of the equations of wave functions~\eqref{eq2.16} and the
respective continuity of currents at all  interfaces of nano-systems
\begin{equation}
\label{eq2.17} \Psi _{\pm 1}^{(p)} \left( {z_{p}}  \right) = \Psi
_{\pm 1}^{(p + 1)} \left( {z_{p}}  \right), \qquad {\left.
{{\frac{{\rd\Psi _{\pm 1}^{(p)} \left( {z} \right)}}{{m_{0(1)} \rd
z}}}} \right|}_{z = z_{p}}  = {\left. {{\frac{{\rd\Psi _{\pm
1}^{(p + 1)} \left( {z} \right)}}{{m_{1(0)} \rd z}}}} \right|}_{z
= z_{p}}  \qquad  (p = 0,\ 1,\ 2,\ 3)
\end{equation}
lead to the system of eight inhomogeneous equations determining
all eight unknown coefficients $B_{\pm} ^{(0)}$, $A_{\pm} ^{(4)}$,
$B_{\pm} ^{(p)}$, $A_{\pm }^{(p)}$ $(p = 1,\ 2,\ 3)$. Thus, now
$\Psi _{\pm} (z)$ functions, the first order correction~-- $\Psi
_{1} (z,t)$ and, consequently, the whole $\Psi (z,t)$ wave
function are completely defined.

According to the quantum mechanics, the density of current of
uncoupling electrons with concentration $n_{{\rm 0}}$ is given by
the formula
\begin{equation}
\label{eq2.18} j(z,t) = {\frac{{\ri e\hbar {\rm} n_{0}}
}{{2m(z)}}}{\left[ {\Psi (z,t){\frac{{\partial} }{{\partial
z}}}\Psi ^{\ast} (z,t) - \Psi ^{\ast }(z,t){\frac{{\partial}
}{{\partial z}}}\Psi (z,t)} \right]}.
\end{equation}
Taking into account the small sizes of two-barrier RTS compared
to the electromagnetic wavelength, in quasi-classic
approximation~\cite{11,12,13,14,15,16,17} the
calculation of the guided current density is performed, determining the real part
of nano-system conductivity
\begin{equation}
\label{eq2.19}
\begin{array}{c} \sigma (\omega ) = \sigma ^{ +} (\omega ) + \sigma ^{
-} (\omega ) = {\displaystyle \frac{{\hbar ^{2}\omega {\rm} n_{0}}
}{{2z_{3} m_{0} \epsilon^{2}}}}{\left[ {k_{ + } \left( {{\left|
{B_{ +} ^{(0)}} \right|}^{2} + {\left| {A_{ +} ^{(4)}}
\right|}^{2}} \right) - k_{ -}  \left( {{\left| {B_{ -} ^{(0)}}
\right|}^{2} + {\left| {A_{ -} ^{(4)}}  \right|}^{2}} \right)}
\right]}.
\end{array}
\end{equation}

Here $\sigma^{+}(\omega )$, $\sigma^{-}(\omega )$ are the
components of conductivity throughout the nano-system and in the
opposite direction, respectively, caused by the electronic currents
from the nano-system after the interaction with
electromagnetic field therein.

\section{Discussion of the results}

The operating characteristics of a separate nano-laser,
nano-detector or quantum cascade nano-devices at their base are
determined by the properties of active conductivity, depending on
spectral parameters (resonance energies and widths) of electron
QSSs. The latter characteristics, in turn, are defined by material
and geometric parameters of a nano-system. Thus, one has to study,
first of all, the spectral parameters of electron QSSs for
further analyzing and understanding the main properties of the active
conductivity. For example, the paper studies the widely
experimentally investigated~\cite{1,2,3,7,8,9,10} plane
two-barrier RTS (figure~\ref{Fig1}), consisting of
In$_{0.53}$Ga$_{0.47}$As wells ($m_{0} = 0.046~m_{e}$) and
In$_{0.52}$Al$_{0.48}$As barriers ($m_{1}= 0.089~m_{e}$). The
difference of electron potential energy between the barrier and
the well is $U = 516$~meV. This structure well satisfies the
developed theory conditions due to the close magnitudes of lattice
constants ($a_{0}= 0.5867$~nm, $a_{1}= 0.5868$~nm), wells
($\varepsilon_{0}= 14.2$) and barriers ($\varepsilon _{1} = 12.7$)
dielectric constants.

\begin{wrapfigure}{i}{0.49\textwidth}
\centerline{\includegraphics[width=0.49\textwidth] {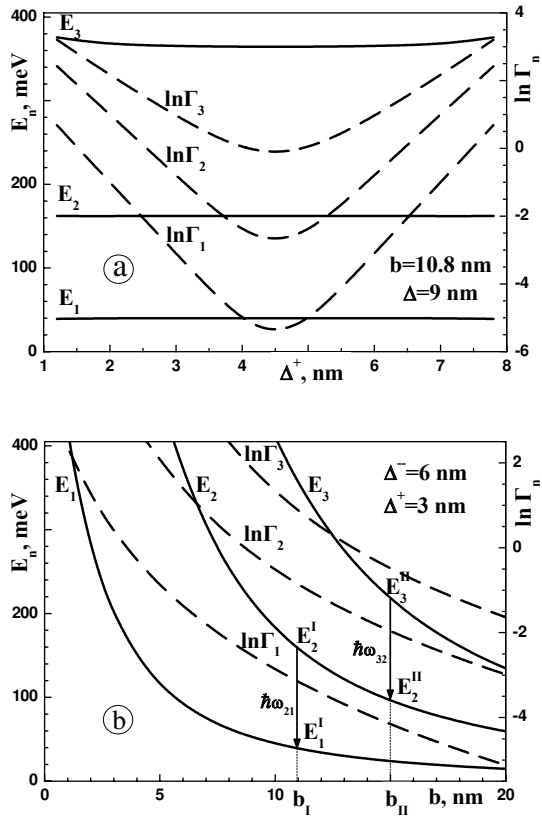}}
\caption{Dependencies of resonance energies $E_{n}$ and logarithms
of resonance widths $\ln\Gamma_{n}$\,: (a) on output barrier width
$\Delta^{+}$ at $\Delta=9$~nm and (b) on well width~$b$.}
\label{Fig2}
\end{wrapfigure}
%
In figure~\ref{Fig2} there are shown the resonance energies
($E_{n}$) and logarithms of resonance widths ($\ln\Gamma_{n}$\,,
in units $\Gamma_{0}= 1$~meV) of the three lowest electron QSSs
depending on geometric parameters of two-barrier RTS: i.e., on the width
of the output barrier $\Delta^{+}$  (fi\-gure~\ref{Fig2}~(a)) at
constant well width $b=10.8$~nm with sum barriers width
$\Delta=\Delta^{-}+\Delta^{+}=9$~nm; potential well width $b$
 (figure~\ref{Fig2}~(b)) at the fixed barriers widths
$\Delta^{-}=6$~nm, $\Delta^{+}=3$~nm. The resonance energy
($E_{n}$) of electron n-th QSS is fixed by the position of maximum
of n-th peak of permeability coefficient and the resonance width
($\Gamma_{n}$)~-- as the width of the same peak at the half of its
height in energy scale.

From figure~\ref{Fig2} it is clear that the change of both
widths of barriers ($\Delta^{-}$, $\Delta^{+}$) almost does not
change the spectrum of resonance energies ($E_{n}$), as square
function of quantum number $n$. That is: $E_{n}=E_{1}n^{2}$, like
the spectrum of quasi-particle in infinitely deep potential well.
Here $E_{1}$ is the resonance energy of the first electron QSS in
two-barrier RTS (fixed, as it was mentioned above, by the position
of the first maximum of permeability coefficient in energy scale).

Contrary to the resonance energies ($E_{n}$), the
magnitudes of resonance widths ($\Gamma_{n}$) essentially depend
on the ratio between the barrier widths. It is clear
(figure~\ref{Fig2}~(a)) that the increasing width of the output
barrier ($\Delta^{+}$) and the respective decreasing width of the
input barrier ($\Delta^{-}=\Delta -\Delta^{+}$) causes a linear
decrease of all $\ln\Gamma_{n}$\,, approaching the minima
magnitudes at $\Delta^{+}=\Delta^{-}=4.5$~nm. At further
$\Delta^{+}$ increase, the magnitudes of $\ln\Gamma_{n}$ linearly
increase. Such a behavior of the width of electron QSSs
($\Gamma_{n}$) in $0<\Delta^{\pm}<\Delta/2$ range can be described
by the typical, for the quantum-barrier systems, dependence:
$\Gamma_{n}=\Gamma_{n}^{0}\exp(-\gamma_{n}\Delta^{\pm})$, where
$\Gamma_{n}^{0}$ is the width of $n$-th virtual QSS, existing at
$\Delta^{\pm}=0$, $\Delta ^{\mp}  = \Delta \ne 0$, $b \ne 0$. The
magnitudes $\gamma_{n}$ characterize the speed of resonance widths
decrease when the barriers width ($\Delta^{\pm}$) increases.

The dependence of electron QSSs spectral parameters ($E_{n}$\,,
$\Gamma_{n}$) on the well width ($b$) is shown in
figure~\ref{Fig2}~(b). In the figure one can see that for the
increasing well width, the resonance energies ($E_{n}$) and
logarithms of resonance widths ($\ln\Gamma_{n}$) shift into the
region of smaller energies. Herein, $E_{n}\sim b^{-3/2}$,  contrary to the energy spectrum in infinitely deep potential well,
where $E_{n} \sim b^{ - 2}$. The decreasing character of resonance
widths is caused by a decrease of resonance energies which is
equivalent to the increase of effective potential barriers above
them.

The numerical calculations prove that the established properties
of spectral parameters ($E_{n}$\,, $\Gamma_{n}$) of electron QSSs
are equitable at any magnitudes of two-barrier RTS geometric
parameters ($b$, $\Delta ^{\pm}$, $\Delta$). Further,  the active conductivity of nano-system
is studied as a function of
energy ($E$) of mono-energetic electronic beam impinging upon
two-barrier RTS and electromagnetic field frequency ($\omega$).
The concentration of electronic current is assumed to be small ($n_{0}
\approx 10^{16}$~cm$^{-3}$), which  allows us to neglect the coupling
between electrons.

In order to establish the optimal geometric design of two-barrier
RTS, when the system operates in the required range of
electromagnetic wave frequencies as separate nano-detector or
nano-laser, one has to analyze the behavior of active conductivity
($\sigma$) together with its components ($\sigma^{-},\
\sigma^{+}$) formed by the input and output currents, depending on
the nano-system geometric parameters. It is clear that the
effectiveness of any nano-device operation would be better at the
maximal absolute value of conductivity ($\sigma$) at the demanded
condition $|\sigma^{+}|\gg|\sigma^{-}|$ and minimal life time
($\tau _{n} = \hbar/\Gamma_{n}$) of electrons in the operating QSSs
(minimizing the negative effect of dissipative processes).

Now let us study the two-barrier RTS conductivity, considering
that the mono-energetic beam of uncoupling electrons impinges on it
from the left, in the direction perpendicular to the planes of the layers.
The  energy of electrons is the same as the resonance energy ($E_{n}
$) of n-th QSS. The electronic current interacts with the
electromagnetic field in such a way that the quantum transitions
take place in the nano-system. As a result,  the
active conductivity is formed, being, as it is well known~\cite{17},
positive (detector) for the transitions accompanied by the
absorption of electromagnetic field energy, and negative (laser)
with the energy radiation.

The numeric calculations prove that the active conductivity is
mainly formed by quantum transitions between two nearest electron
QSSs, because the transitions between the states with equal parity
are forbidden and the transitions into the far resonance states
are weak. Therefore, studying the conductivity and its components,
we consider only the transitions between the nearest states.

In order to detect (or radiate) the electromagnetic field with
frequency $\omega$ due to the quantum transitions between two QSSs
with the resonance energies $E_{n}$ and $E_{n\pm 1}$\,, it is
necessary that the energy ($E$) of electrons impinging upon the system should
correspond to the energy ($E_{n}$) of those $n$-th state from
which the transition occurs. Then, the electromagnetic field energy
is defined as: $\hbar \omega _{n,n\pm 1} = |E_{n}-E_{n\pm 1}| =
(2n\pm 1) E_{1} $\,. Consequently, the estimation of electron
ground QSS energy is $E_{1} = \hbar \omega _{n,n\pm 1}/(2n\pm 1)$
and thus, the energy of the impinging electronic current is: $E_{n}=
\hbar \omega _{n,n\pm 1} n^{2} / (2n\pm 1)$. The latter, due to
the established properties (figure~\ref{Fig2}~(a),~(b)), weakly
depends on  the barrier widths ($\Delta^{-}$, $\Delta^{+}$) and is
mainly determined by the well width ($b$) of two-barrier RTS.

Thus, when the energy ($\hbar\omega$) of the detected (or
radiated) electromagnetic field is known, first of all the
estimation of the required well width $b_{0}=\pi\hbar n(2m_{0}
E_{1} )^{-1/2} = (\hbar \pi ^{2}(2n\pm 1)/2m_{0}\omega
_{n{n}'})^{1/2}$ can be obtained within the model of infinitely
deep potential well. Then, a more exact magnitude of the
two-barrier RTS well width ($b$) is found as its variation in
$b_{0}$ vicinity.

\begin{figure}[!b]
\centerline{\includegraphics[width=0.9\textwidth]{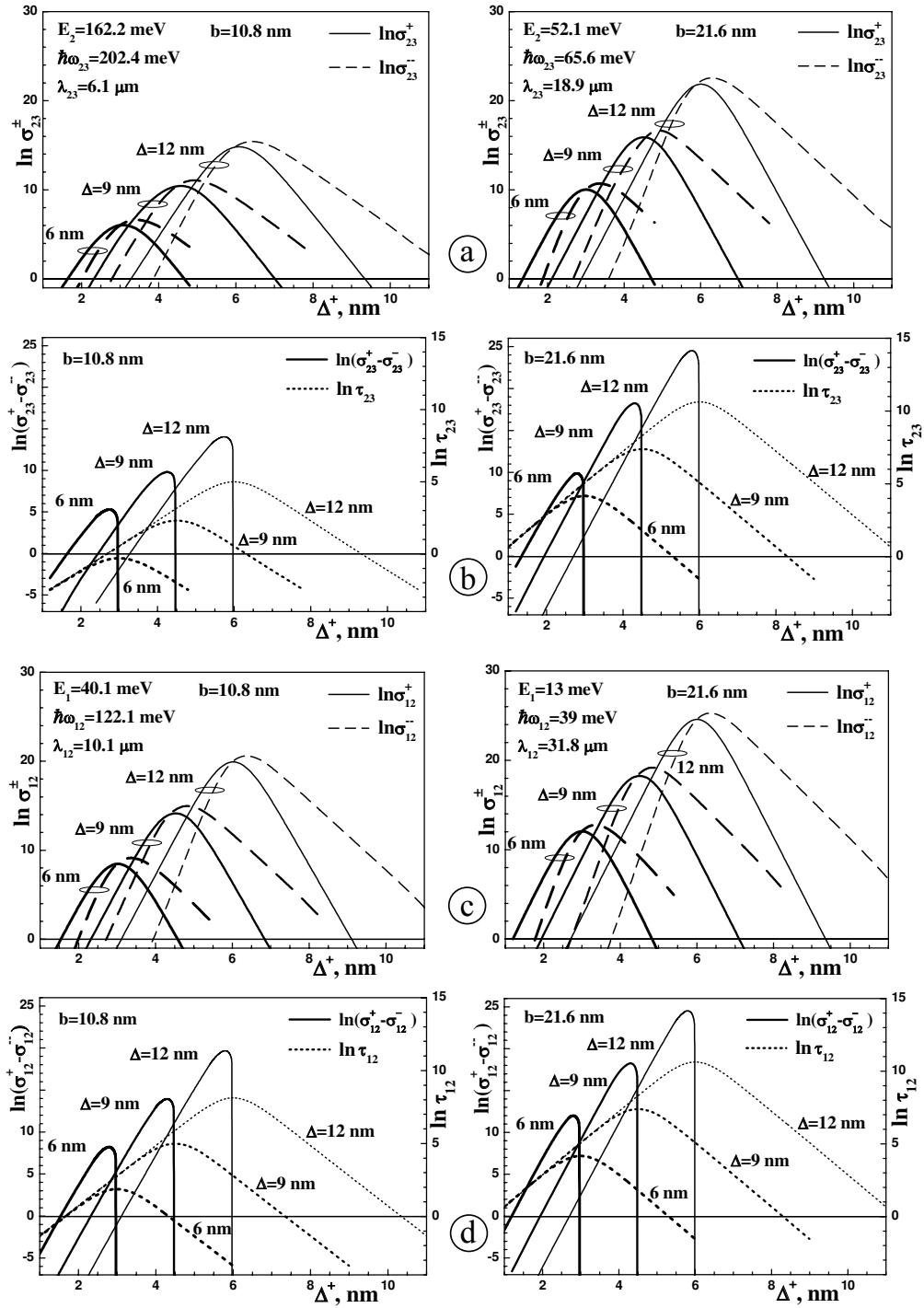}}
\caption{Dependencies of conductivity logarithms
$\ln\sigma_{23}^{+}$\,, $\ln \sigma_{23}^{-}$\,,
$\ln\sigma_{12}^{+}$\,, $\ln\sigma_{12}^{-}$\,,
$\ln(\sigma_{23}^{+}-\sigma_{23}^{-})$, $\ln
(\sigma_{12}^{+}-\sigma_{12}^{-})$ and life times $\ln \tau _{23}
$\,, $\ln\tau_{12}$ on output barrier width $\Delta^{+}$ at
different well widths $b$ and sum $\Delta$, for the two-barrier
RTS operating as detector.} \label{Fig3}
\end{figure}

\begin{figure}[!h]
\centerline{\includegraphics[width=0.9\textwidth]{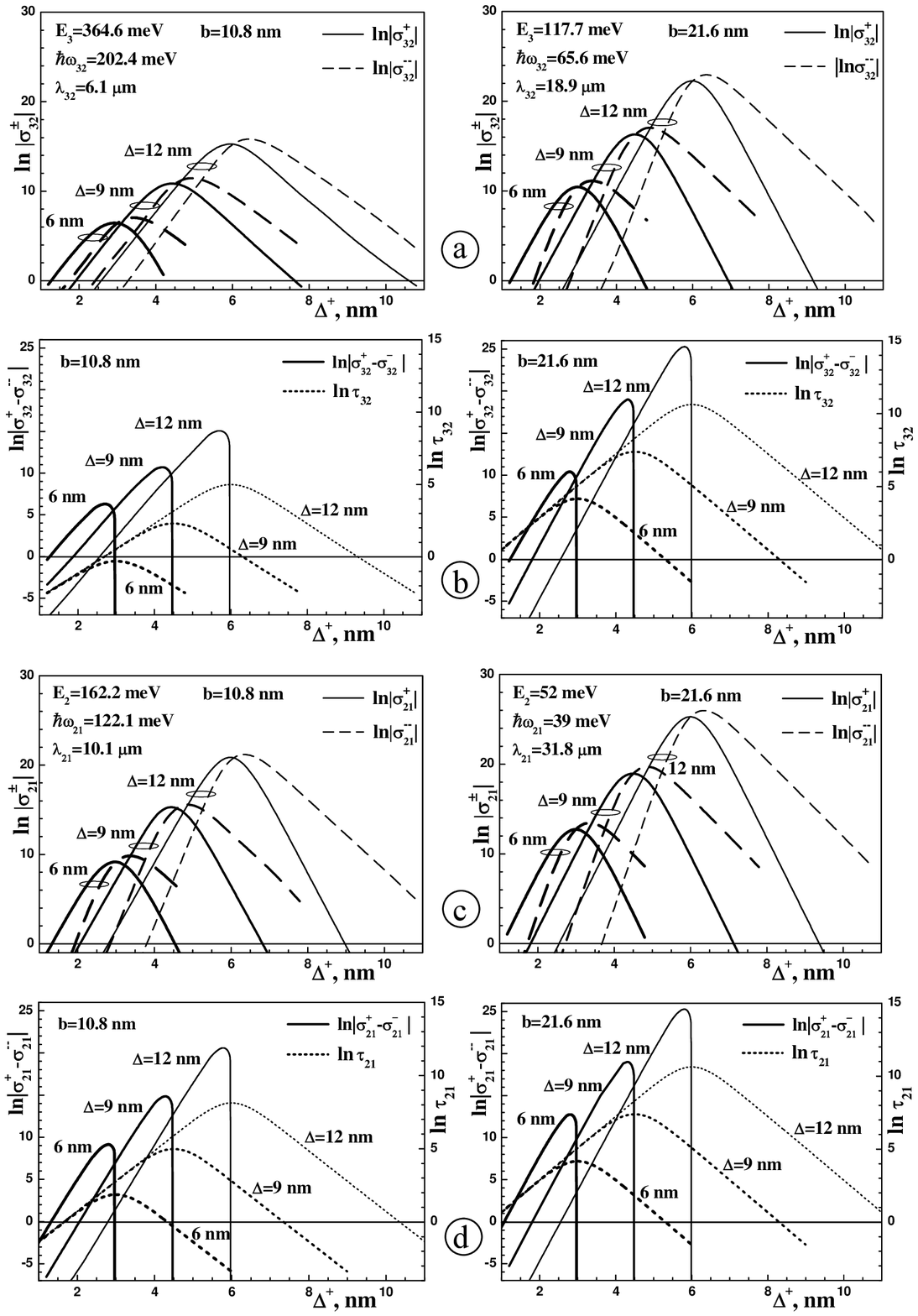}}
\caption{Dependencies of conductivity logarithms
$\ln|\sigma_{32}^{+}|$\,, $\ln|\sigma_{32}^{-}|$\,,
$\ln|\sigma_{21}^{+}|$\,, $\ln| \sigma_{21}^{-}|$\,,
$\ln|\sigma_{32}^{+}-\sigma_{32}^{-}|$\,,
$\ln|\sigma_{21}^{+}-\sigma_{21}^{-}|$ and life times
$\ln\tau_{32}$\,, $\ln\tau_{21}$ on output barrier width
$\Delta^{+}$ at different well width $b$ and sum $\Delta$, for the
two-barrier RTS operating as laser.} \label{Fig4}
\end{figure}

The developed theory for two-barrier RTS active conductivity is
valid for any type of quantum transitions (laser or detector).
Further, to optimize the design of the system under research as an
active element of nano-laser or nano-detector, there is performed the
calculation of electron QSSs life times and maxima of active
conductivity together with its components ($\sigma^{-}$,
$\sigma^{+}$). The latter are formed by detector transitions with
absorption (figure~\ref{Fig3}) and by laser transitions with
radiation (figure~\ref{Fig4}) of electromagnetic field energy
($\hbar\omega$) with the respective wave-lengths ($\lambda$) shown
in figures~\ref{Fig3}~(a),~(c) and \ref{Fig4}~(a),~(c). The widths of the wells: $b =
10.8$~nm, $21.6$~nm and the widths of the barriers:
$\Delta=\Delta^{-}+\Delta^{+} = 6$~nm, 9~nm, 12~nm were used as
typical ones~\cite{3,7,8,9,10} for the experimentally investigated QCDs
or QCLs with the range of operating frequencies being in the
sub-infrared windows of atmosphere transparency.

The calculated conductivity logarithms: forward $\ln |\sigma
_{12(21)}^{+}|$, $\ln |\sigma_{23(32)}^{ +} |$ and backward $\ln |
\sigma_{12(21)}^{-}|$, $\ln |\sigma_{23(32)}^{-}|$ and their
differences $\ln|\sigma_{12(21)}^{+}-\sigma_{12(21)}^{-}|$,
$\ln|\sigma_{23(32)}^{+}-\sigma_{23(32)}^{-}|$ (in units
$\sigma_{0}= 1$~S/cm) formed by the detector ($1 \to 2$, $2 \to
3$) and laser ($2 \to 1$, $3 \to 2$) quantum transitions; and
logarithms of electron life times sum in operating QSSs $\ln \tau
_{n,n\pm 1} $\,, where $\tau _{n,n\pm 1} = \tau _{n} + \tau _{n\pm
1}$ (in units $\tau _{0} = 1 $~s) are presented in
figures~\ref{Fig3}, \ref{Fig4} respectively, as functions of output
barrier width $\Delta^{+}$ at the condition of the constant sum of
barriers widths: $\Delta =\Delta^{-}+\Delta^{+}=\rm const$ (6~nm,
9~nm, 12~nm).

In figures~\ref{Fig3}~(a),~(c) and \ref{Fig4}~(a),~(c) one can see that
independently of the well width ($b$) and sum of barriers widths
($\Delta$) all $\ln|\sigma_{12}^{\pm}|$, $\ln|\sigma_{23}^{\pm}|$
and $\ln|\sigma_{21}^{\pm}|$, $\ln|\sigma_{32}^{\pm}|$ magnitudes
qualitatively similarly depend on the output barrier width
($\Delta^{+}$). Herein, it is evident that $\ln|\sigma_{21}^{\pm}|
> \ln |\sigma_{12}^{\pm}| > \ln|\sigma_{32}^{\pm}| > \ln|\sigma_{23}^{\pm}|$. Their main
properties are as follows.

The increase of $\Delta^{+}$ (and the  corresponding decrease of
$\Delta^{-}$) in the range $0<\Delta^{+}<\Delta/2$ causes the linear
increase of $\ln|\sigma_{12(21)}^{+}|$, $\ln
|\sigma_{23(32)}^{+}|$ and $\ln|\sigma_{12(21)}^{-}|$,
$\ln|\sigma_{23(32)}^{-}|$ magnitudes. Herein, there are always
satisfied the inequalities:
$\ln|\sigma_{12(21)}^{+}|>\ln|\sigma_{12(21)}^{-}|$,
$\ln|\sigma_{23(32)}^{+}|>\ln|\sigma_{23(32)}^{-}|$. In the range
of output barrier widths $\Delta/2<\Delta^{+}<\Delta$, on the
contrary, these magnitudes are linearly decreasing with
$\ln|\sigma_{12(21)}^{-}|>\ln|\sigma_{12(21)}^{+}|$,
$\ln|\sigma_{23(32)}^{-}|>\ln|\sigma_{23(32)}^{+}|$. In the
symmetric two-barrier RTS, when $\Delta^{+}=\Delta^{-}=\Delta/2$,
we have $\ln| \sigma_{12(21)}^{+}|=\ln|\sigma_{12(21)}^{-}|$,
$\ln|\sigma_{23(32)}^{+}|=\ln|\sigma_{23(32)}^{-}|$ which is
completely understandable from physical considerations because the
electronic current, after the interaction with the electromagnetic
field, flows from the two-barrier RTS by two currents equal over
magnitudes and opposite over directions.

The established properties of active conductivity ($\sigma$) and
its components ($\sigma_{12(21)}^{+}$, $\sigma_{12(21)}^{-}$ and
$\sigma_{23(32)}^{+}$, $\sigma_{23(32)}^{-}$) lead to the evident
conclusion. Now it is clear that the condition
$|\sigma_{12(21)}^{+}|\gg|\sigma_{12(21)}^{-}|$ or
$|\sigma_{23(32)}^{+}|\gg|\sigma_{23(32)}^{-}|$, at which the
two-barrier RTS optimally operates as detector or laser, is
fulfilled only in the interval of widths $\Delta^{+}<\Delta/2$
($\Delta/2<\Delta^{-}=\Delta-\Delta^{+}$), i.e., at the ascending
parcels of $\ln\sigma_{12(21)}^{+}$\,, $\ln\sigma_{23(32)}^{+}$
and $\ln\sigma_{12(21)}^{-}$\,, $\ln\sigma_{23(32)}^{-}$ as
functions of $\Delta^{+}$. There is performed a calculation of
$\ln|\sigma _{12(21)}^{+}-\sigma_{12(21)}^{-}|$, $\ln
|\sigma_{23(32)}^{+}-\sigma_{23(32)}^{-}|$ and $\ln \tau _{12}$\,,
$\ln\tau_{23}$ as functions of $\Delta^{+}$ in order to establish
the best ratio between $\Delta^{+}$ and $\Delta^{-}$ magnitudes
(at fixed $\Delta=6$~nm, 9~nm, 12~nm). The results are shown in
figures~\ref{Fig3}~(b),~(d) and \ref{Fig4}~(b),~(d).

From figures~\ref{Fig3} and~\ref{Fig4} one can see that at any
$\Delta$, maximum magnitudes of
$\ln|\sigma_{12(21)}^{+}-\sigma_{12(21)}^{-}|$, $\ln |\sigma
_{23(32)}^{+}-\sigma_{23(32)}^{-}|$ are approached at such widths
$\Delta^{+}=\Delta_{m}^{+}$\,, which are in the small vicinity
from the left hand side of $\Delta ^{+}=\Delta/2$. It is clear that as
far as the output barrier width $\Delta_{m} ^{+}$ is only a little
bit smaller than the input barrier width ($\Delta_{m}
^{-}=\Delta-\Delta_{m} ^{+}$), hence, the forward electronic
current through the thinner output barrier is much bigger than the
backward current through the thicker input barrier, observed in
figure~\ref{Fig3}~(b),~(d) and \ref{Fig4}~(b),~(d).

For the optimal design of two-barrier RTF and coherent character
of electromagnetic wave (in case of laser), the evident physical
requirements must be fulfilled: the life time ($\tau_{n,n\pm 1}$) of
electron in both operating QSSs should  not be bigger than the time
($\tau_{\rm d}$) of dissipative processes (interaction with
phonons, impurities and so on). According to the known estimations
[3] $\tau_{\rm d}\approx 20$~s or $\ln\tau_{\rm d}\approx 3$. Now,
there are obtained the consequent estimations of optimal widths of
both barriers $\Delta^{+}$ and $\Delta^{-}$ at which, at the
conditions $|\sigma_{12(21)}^{+}|\gg|\sigma_{12(21)}^{-}|$,
$|\sigma_{23(32)}^{+}|\gg|\sigma_{23(32)}^{-}|$, the magnitudes
$\sigma_{12(21)}$\,, $\sigma_{23(32)}$ are maximum due to the
linear dependence of
$\ln|\sigma_{12(21)}^{+}-\sigma_{12(21)}^{-}|$,
$\ln|\sigma_{23(32)}^{+}-\sigma_{23(32)}^{-}|$ and $\ln|
\sigma_{12(21)}^{\pm}|$, $\ln|\sigma_{23(32)}^{\pm}|$ maxima on
$\Delta$ (figure~\ref{Fig3}, \ref{Fig4}) at fixed well width ($b$).
The magnitudes of all evaluated characteristics are shown in
figures~\ref{Fig3}, \ref{Fig4}. From the figure one can see that
when the well width ($b$) increases, the optimal sum of both
barriers widths ($\Delta=\Delta^{+}+\Delta^{-}$), confined by
dissipative processes scattering time ($\tau_{\rm d}$)\,,
decreases.

The detection or radiation of electromagnetic field with certain
frequency can be obtained due to quantum transitions between QSSs
in different two-barrier RTSs with respective widths of potential
wells. There arises a question: which one of the two RTSs is more
optimal~-- the one with smaller well width and operating at the
transitions between lower levels or the one with bigger well width
and operating at the respective transitions between higher levels.

There is an important property of two-barrier RTS, which is really clear
from physical considerations. When the electron energy is equal to
the resonance energy of any level except the ground one, the
quantum transitions into the lower QSSs~(figure~\ref{Fig4}) are
more probable than into the higher ones~(\ref{Fig3}). Consequently, the
nano-system has a negative conductivity and operates in the regime
of electromagnetic field radiation. Thus, in a detector regime, the
two-barrier RTS operates only  when the energy of
electronic current is equal to the resonance energy of ground QSS
and quantum transition into the second state occurs with the
absorption of electromagnetic field energy. We should note that all the
known experimentally utilized nano-detectors operate exactly  in this
regime~\cite{7,8,9,10}.

As far as the laser regime is concerned, this is realized within the
quantum transitions between the neighbouring levels from arbitrary higher
into arbitrary lower states. For example, we can compare the
advantages and disadvantages of two different RTS, radiating the
electromagnetic field in one of the ranges of atmosphere
transparency windows ($\lambda = 8 - 14~\mu\rm m$ or $\hbar \omega
= 89 - 155$~meV). It is clear from figure~\ref{Fig2}~(b) that the
radiation with the field energy $\hbar \omega = 122$~meV can be
realized by two two-barrier RTSs with equal barrier widths
($\Delta^{-} = 6$~nm, $\Delta^{+}= 3$~nm) but different well
widths: I) at the transition $2 \to 1$ ($b_{I}=10.8$~nm)
$\tau_{1}^{I}= 14.6$~ps, $\tau_{2}^{I} = 1.5$~ps,
$\tau_{21}^{I}=\tau_{2}^{I}+\tau_{1}^{I}=16.1$~ps, $E_{2}^{I}=
162$~meV, $|\sigma_{21}^{+}|= 1200$~S/cm,
$|\sigma_{21}^{-}|=1$~S/cm; II) at the transition $3 \to 2$
($b_{II}=15$~nm) $\tau_{2}^{II}= 4.7$~ps, $\tau_{3}^{II}= 1.1$~ps,
$\tau_{32}^{II}=\tau_{3}^{II}+\tau_{2}^{II}=5.8$~ps,
$E_{2}^{II}=219$~meV, $|\sigma_{32}^{+}|= 811$~S/cm,
$|\sigma_{21}^{-}|=2$~S/cm. Comparing the both nano-systems, one
can see, even if the life-time ($\tau_{12}^{I}$) of electron in
the operating QSS for the system I is almost three times bigger than the
life time ($\tau_{32}^{II}$) for the system II, but, herein, the
starting electron current ($j_{0}^{I}\sim\sqrt {E_{2}^{I}}$) at
system I is 1.2 times smaller than the one ($j_{0}^{II}\sim \sqrt
{E_{3}^{II}}$) at the system II. The intensity of radiation,
proportional to the conductivity, for the system I is about 1.5
times bigger than for the II system. Thus, the increase of QSSs
life times for the system I is not bigger than the scattering
times of electrons due to the dissipative processes (phonons,
impurities and so on) destroying the coherence. Thus, the optimal
two-barrier RTS is the one operating at quantum transitions
between two lowest QSSs.

Finally, we should note that the theory for the active conductivity
of two-barrier RTS developed within the framework of simple rectangular
potentials model and different electron effective masses
approximation can be used in future as a basis of investigation
and optimization of complicated open RTS with bigger number of
wells and barriers, intensively utilized as operating elements of
quantum cascade nano-lasers and nano-detectors.


\ukrainianpart

\title{╥хюЁ│  ръЄштэю┐ хыхъЄЁюээю┐ яЁют│фэюёЄ│ яыюёъю┐ фтюсрЁ'║Ёэю┐ эрэюёшёЄхьш  ъ Ёюсюўюую хыхьхэЄр ътрэЄютюую ърёърфэюую ырчхЁр ўш фхЄхъЄюЁр}
\author{╠.┬.~╥ърў, ▐.╬.~╤хЄ│, ┬.╬.~╠рЄ│║ъ, ╬.╠.~┬ющЎхї│тё№ър}
\address{╫хЁэ│тхЎ№ъшщ эрЎ│юэры№эшщ єэ│тхЁёшЄхЄ │ьхэ│ ▐Ё│  ╘хф№ъютшўр, \\58012 ╫хЁэ│тЎ│, тєы. ╩юЎ■сшэё№ъюую 2}

\makeukrtitle

\begin{abstract}
\tolerance=3000%
╙ ьюфхы│ яЁ ьюъєЄэшї яюЄхэЎ│ры│т │ Ё│чэшї хЇхъЄштэшї ьрё хыхъЄЁюэр
т Ё│чэшї хыхьхэЄрї яыюёъю┐ фтюсрЁ'║Ёэю┐ Ёхчюэрэёэю-Єєэхы№эю┐
ёЄЁєъЄєЁш (─┴╨╥╤) ЁючтшэєЄр ътрэЄютю-ьхїрэ│ўэр ЄхюЁ│  ёяхъЄ\-Ёры№\-эшї
ярЁрьхЄЁ│т ътрч│ёЄрЎ│юэрЁэшї ёЄрэ│т │ яЁют│фэюёЄ│ Ў│║┐ ёшёЄхьш фы 
тшярфъє ью\-эю\-хэхЁ\-ух\-Єшў\-эю\-ую яєўър хыхъЄЁюэ│т,  ъ│ тчр║ьюф│■Є№ ч
хыхъЄЁюьруэ│Єэшь яюыхь.
╧юърчрэю, ∙ю эрэю-─┴╨╥╤ ьюцх ёыєуєтрЄш юъЁхьшь хыхьхэЄюь рсю
ръЄштэшь хыхьхэЄюь ърё\-ърф\-эю\-ую ырчхЁр ўш фхЄхъЄюЁр. ═р яЁшъырф│
хъёяхЁшьхэЄры№эю фюёы│фцєтрэю┐ эрэюёшёЄхьш
In$_{0.53}$Ga$_{0.47}$As/In$_{0.52}$Al$_{0.48}$As яюърчрэю, ∙ю є
фхЄхъЄюЁэюьє │ ырчхЁэюьє Ёхцшьрї ЁюсюЄр ─┴╨╥╤ ║ юяЄшьры№эю■ (ч
эрщс│ы№°ю■ яЁют│фэ│ёЄ■ яЁш эрщьхэ°юьє ёЄЁєь│ чсєфцхээ ), ъюыш тюэр
яЁрЎ■║ эр ътрэЄютшї яхЁхїюфрї ь│ц эрщэшцўшьш ътрч│ёЄрЎ│юэрЁэшьш
ёЄрэрьш.
\keywords Ёхчюэрэёэю-Єєэхы№эр ёЄЁєъЄєЁр, яЁют│фэ│ёЄ№, ътрэЄютшщ
ырчхЁ, ътрэЄютшщ фхЄхъЄюЁ

\end{abstract}

\end{document}